\documentclass[a4paper]{article}
\usepackage{INTERSPEECH2020}

\usepackage{multirow}
\usepackage{enumitem}
\usepackage{hyperref}
\newcommand*\samethanks[1][\value{footnote}]{\footnotemark[#1]}

\title{Attentron: Few-Shot Text-to-Speech Utilizing Attention-Based Variable-Length Embedding}

\name{Seungwoo Choi\sthanks{\quad Equal contributions, listed in alphabetical order.}, Seungju Han\samethanks[1], Dongyoung Kim\samethanks[1], Sungjoo Ha\sthanks{\quad Corresponding author.}}
\address{Hyperconnect, Seoul, South Korea}
\email{\{seungwoo.choi, seungju.han, dongyoung.kim, shurain\}@hpcnt.com}

\begin{document}
\maketitle

\begin{abstract}

On account of growing demands for personalization, the need for a so-called few-shot TTS system that clones speakers with only a few data is emerging.
To address this issue, we propose Attentron, a few-shot TTS model that clones voices of speakers unseen during training.
It introduces two special encoders, each serving different purposes.
A fine-grained encoder extracts variable-length style information via an attention mechanism, and a coarse-grained encoder greatly stabilizes the speech synthesis, circumventing unintelligible gibberish even for synthesizing speech of unseen speakers.
In addition, the model can scale out to an arbitrary number of reference audios to improve the quality of the synthesized speech.
According to our experiments, including a human evaluation, the proposed model significantly outperforms state-of-the-art models when generating speech for unseen speakers in terms of speaker similarity and quality.
\end{abstract}


\section{Introduction}

Recent successes of deep learning methods for speech synthesis enabled text-to-speech (TTS) systems to synthesize realistic and natural speech~\cite{skerry2018towards, ping2018clarinet, shen2018natural}.
Beyond this capability, there have been growing demands for personalization, putting pressure on modern TTS systems to generate customized voices of high quality.
Conventional multi-speaker TTS systems~\cite{Park2019MultiSpeakerES, gibiansky2017deep} require a substantial amount of data merely to model the speakers observed during training.
Unfortunately, many personalized applications can only afford a handful of reference data (e.g., restoring communication ability to people who lost their voice).
Such needs call for a speaker cloning system so-called \textit{few-shot TTS} that can function with only a few reference samples.

To implement few-shot TTS, previous studies suggested a speaker adaptation process~\cite{kons2019high, chen2018sample, bollepalli2019lombard, deng2018modeling} that pre-trains models on a large dataset of many speakers and then continues training on a small dataset of target speakers.
Such methods, however, require at least a few minutes of audio samples together with the additional fine-tuning process.
Therefore, it is less attractive under in-the-wild scenarios where immediate voice cloning of arbitrary new speakers cannot be avoided.

Some prior approaches predicted a speaker embedding from speech to clone unseen speakers without fine-tuning, using a speaker encoder jointly trained with the TTS model~\cite{nachmani2018fitting, arik2018neural} or a model individually trained for a speaker verification task~\cite{hu2019neural, cooper2019zero, jia2018transfer}.
However, it is challenging to produce a single embedding that represents every utterance characteristic, including the speaker identity and speaking style.
In fact, it was reported that a single embedding works poorly if the reference speech is shorter than the target speech~\cite{wang2018style}.

To tackle such a problem, previous studies proposed several specialized embeddings, each with a different set of responsibilities to represent diverse speech attributes (e.g., speaking style prosody, and noise)~\cite{wang2018style, stanton2018predicting, hsu2018hierarchical, hsu2019disentangling, battenberg2019effective, chen2019cross} or switched over to a variable-length embedding to maintain temporal information~\cite{lee2019robust, sun2020fully, sun2020generating}.
However, most of them focused not on cloning a target speaker but on controlling the disentangled attributes~\cite{wang2018style, stanton2018predicting, hsu2018hierarchical, hsu2019disentangling, battenberg2019effective, sun2020fully, sun2020generating}, and some of them could not synthesize speech of an unseen speaker due to their architectural limitations~\cite{lee2019robust}.
In addition, there have been few attempts to utilize multiple reference samples to enhance the speech quality~\cite{arik2018neural} even though several utterances of a target speaker are available during inference in a real-world scenario.

In this paper, we propose \textit{Attentron}, a novel architecture of few-shot TTS for unseen speakers.
It consists of a \textit{fine-grained encoder}, which includes an attention mechanism to extract detailed styles from multiple reference samples and a \textit{coarse-grained encoder}, which extracts overall information of speech and helps to stabilize the output.
Our contributions are as follows:

\begin{itemize}[leftmargin=*]
    \item We utilize two speech encoders, the fine-grained encoder and the coarse-grained encoder, to enable multi-speaker TTS to clone unseen target speakers with only a few reference samples.
    The encoders generate a variable-length embedding and a global embedding, respectively, and utilize them to condition the TTS system.
    
    \item We propose an attention mechanism that finds only the relevant positions among the audio frames of the multiple references.
    It allows the model to take any number of the reference samples, and the quality improves with more reference samples.

    \item We compare the proposed model with state-of-the-art methods for multi-speaker TTS that can clone unseen speakers. 
    Our experimental results, including human evaluations, show that the proposed model outperforms the state-of-the-art methods.\footnote {Samples available at \url{https://hyperconnect.github.io/Attentron}.}
\end{itemize}

\section{Attentron Architecture}\label{section:architecture}

\begin{figure}[t]
    \includegraphics[width=1\linewidth]{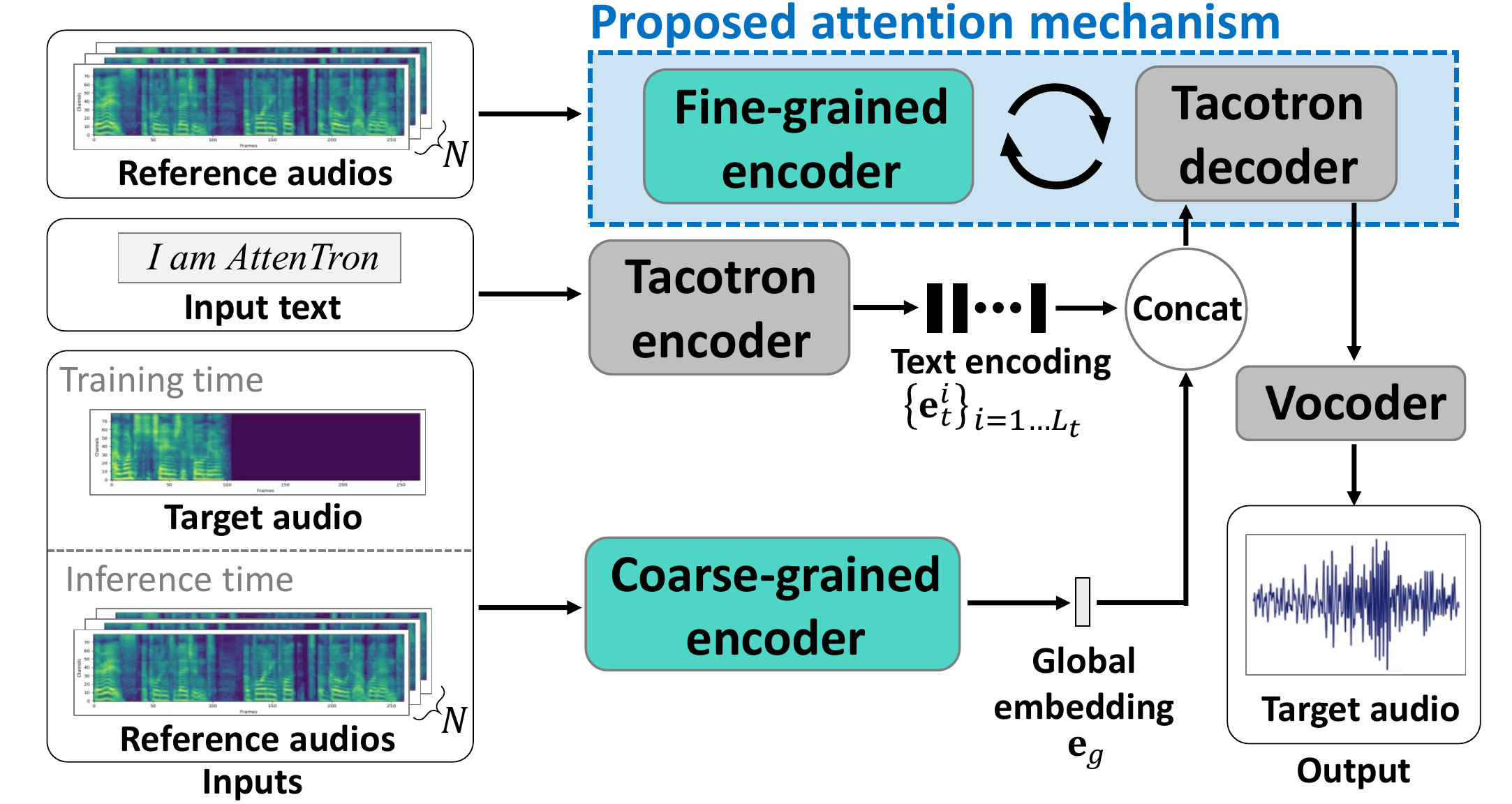}
    \centering
    \caption{
    Overall architecture of Attentron.
    }
    \label{fig:architecture}
\end{figure}

Figure~\ref{fig:architecture} illustrates the overall architecture of the proposed few-shot TTS for unseen speakers, named Attentron.
It is based on Tacotron 2~\cite{shen2018natural}, which takes a text sequence as an input and generates a sequence of mel spectrogram frames as an output.
What differentiates our model is the presence of two additional encoders, the coarse-grained encoder and the fine-grained encoder.
In Attentron, the fine-grained encoder extracts detailed style from multiple (i.e., $\textit{N}$ in Figure~\ref{fig:architecture}) reference audio files, and the coarse-grained encoder extracts overall information of speech, thereby enabling high-quality few-shot TTS for unseen targets.
In our experiments, we use WaveRNN~\cite{kalchbrenner2018efficient} vocoder.

To synthesize speech, first, the coarse-grained encoder generates a \textit{global embedding}, $\mathbf{e}_g$, where a single vector aggregates the temporal dimension of input audio files.
Then, we broadcast-concatenate it with the text encoding, $\{\mathbf{e}^i_t\}_{i=1...L_t}$, generated by the Tacotron encoder.
It forms the input sequence for the Tacotron decoder, $\{(\mathbf{e}^i_t, \mathbf{e}_g)\}_{i=1...L_t}$, where $L_t$ represents the length of the text encoding.
Equipped with the proposed attention mechanism, the fine-grained encoder extracts a \textit{variable-length embedding} maintaining temporal dimension and feeds them to the Tacotron decoder while it synthesizes the spectrogram frames autoregressively.
Finally, the vocoder converts the spectrogram into the audio.

\subsection{Fine-grained Encoder and Attention Mechanism}\label{section:fine_grained_encoder}

\begin{figure}[t]
    \includegraphics[width=1\linewidth]{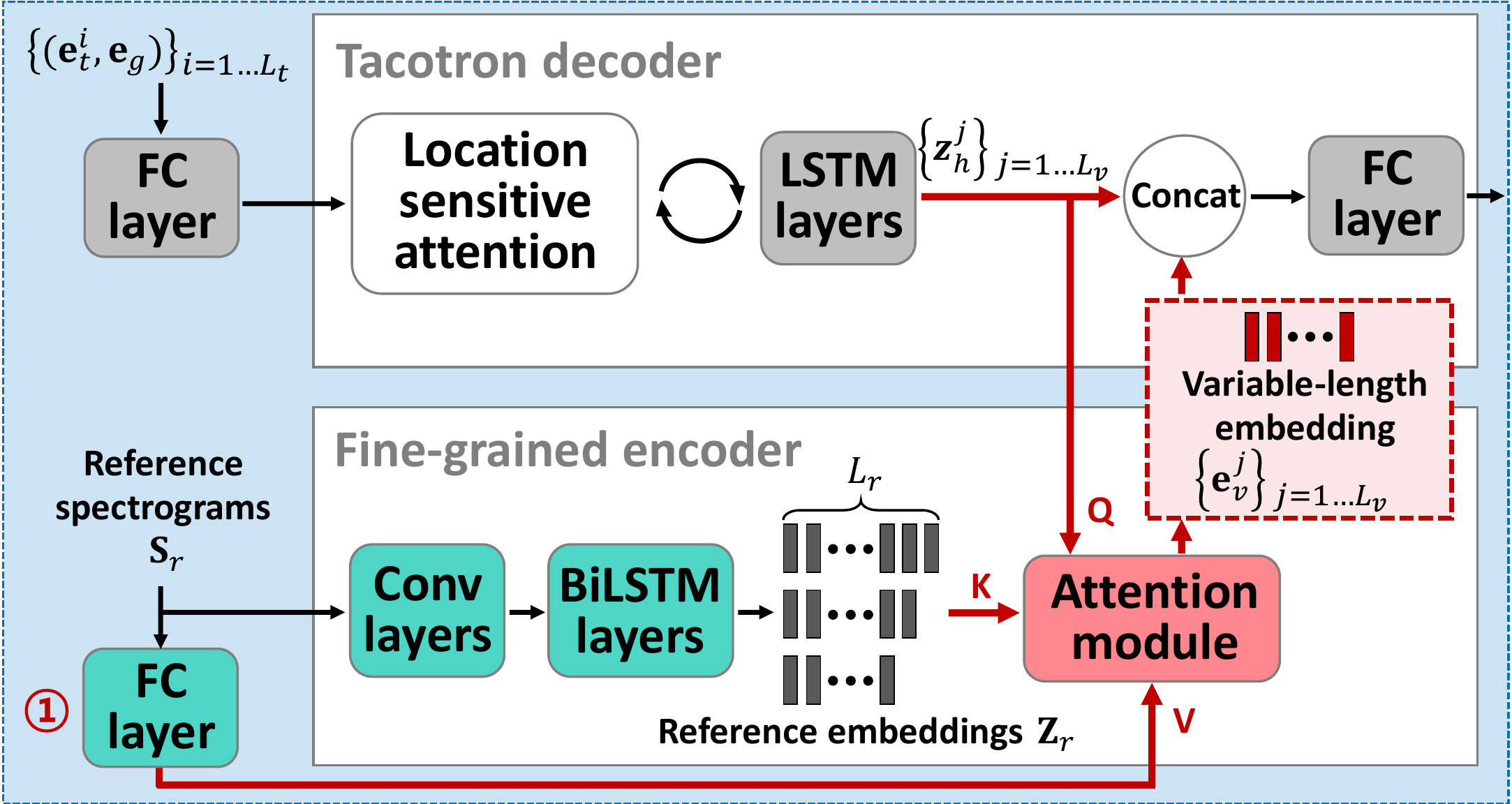}
    \centering
    \caption{
    Details of proposed attention mechanism.
    }
    \label{fig:attention_module}
\end{figure}

Figure~\ref{fig:attention_module} illustrates the composition of the proposed fine-grained encoder and how the attention mechanism works.
It aims to (1) make good use of multiple reference audio files, (2) utilize a variable-length embedding, not just a single global embedding, to maintain detailed information, and (3) leverage features near to raw reference audio for better generalization.
We use scaled dot-product attention~\cite{vaswani2017attention}, which allows the model to attend to the most relevant frames in reference audio spectrograms.

The inputs to the fine-grained encoder are $N$ random reference audio files spoken by the same speaker as the target audio.
We first convert the reference audio files to the mel spectrograms with paddings to match the maximum length of the spectrograms, $L_r$.
Converted reference spectrograms, $\mathbf{S}_r \in \mathbb{R}^{N \times L_r \times n_{mels}}$, are passed to two convolutional layers followed by two bidirectional LSTM layers to give reference embeddings, $\mathbf{Z}_r \in \mathbb{R}^{N \times L_r \times d_r}$, where $n_{mels}$ is the number of mel bins. 
Given the hidden state of decoder LSTM at $j$-th decoding step, $\mathbf{z}_h^j$, the attention query $\mathbf{Q}^j$, key $\mathbf{K}$, value $\mathbf{V}$, and a $j$-th component of variable-length embedding, $\mathbf{e}_v^j \in \mathbb{R}^{d_v}$, are calculated as follows:
\begin{equation*}
\begin{aligned}
    & \mathbf{Q}^j = \mathbf{z}_h^j \mathbf{W}_q & \in & ~\mathbb{R}^{d_m} \\
    & \mathbf{K} = \mathbf{Z}_r \mathbf{W}_k & \in & ~\mathbb{R}^{N \times L_r \times d_m} \\
    & \mathbf{V} = \mathbf{S}_r \mathbf{W}_v & \in & ~\mathbb{R}^{N \times L_r \times d_v}
\end{aligned}
\end{equation*}
\begin{equation*}
    \mathbf{e}_v^j = A(\mathbf{Q}^j, \mathbf{K}, \mathbf{V}) = \text{softmax}\left(\frac{f(\mathbf{Q}^j)f(\mathbf{K})^T}{\sqrt{d_m}}\right)f(\mathbf{V}),
\end{equation*}
where $f : \mathbb{R}^{d_1 \times ... \times d_n \times c} \rightarrow \mathbb{R}^{(d_1 \times ... \times d_n) \times c}$ is a flattening function, and all $\mathbf{W}$ are linear projection matrices.
Each component of variable-length embedding, $\mathbf{e}_v^j$, is concatenated with $\mathbf{z}_h^j$ and fed into fully-connected layer.
The above process iterates autoregressively until generating spectrogram is completed (e.g., it iterates $L_v$ times in Figure~\ref{fig:attention_module}).

Note that the reference spectrograms are passed through only one fully-connected layer to generate the attention values (\textcircled{\raisebox{-0.9pt}{1}} in Figure~\ref{fig:attention_module}).
The intuition is that the more it has following layers, the more prone to overfitting to the speakers in the training data.
Further analysis is described in Section \ref{subsection:ablation}.

\subsection{Coarse-grained Encoder}

Speech from the same speaker has different characteristics, such as emotion and prosody, even with the same transcript.
Thus, synthesizing speech only from the input text may suffer from unstable speech synthesis as it is a non-deterministic \textit{one-to-many problem} by nature.
To stabilize it, the coarse-grained encoder is designed to generate a global embedding, which includes overall information of the target speech.
Giving the outline of the desired output, it narrows down the range of the output speech and makes it close to a \textit{one-to-one problem}.
The previous approaches~\cite{wang2018style, stanton2018predicting, hsu2018hierarchical, hsu2019disentangling, battenberg2019effective, sun2020fully, sun2020generating} aimed to control the speech characteristics of the output speech through an encoder.
Consequently, it was based on complex architecture or accompanied by additional training objectives.
In contrast, our encoder focuses only on stabilizing the synthesis, and, accordingly, we use a relatively simple network generating a global embedding, $\mathbf{e}_g \in \mathbb{R}^{d_g}$, without any additional loss function.

The coarse-grained encoder is based on the encoder architecture proposed in \cite{hsu2018hierarchical}, which has two convolutional layers followed by two bidirectional LSTM layers.
It has an average pooling layer at last to generate a global vector.
Note that we utilize the target audio file as the input during training and utilize reference samples spoken by the target speaker at inference time.
The last layer averages out the embeddings from multiple reference samples to get the single embedding during inference.

\section{Experimental Method}\label{section:method}

\subsection{Experimental Setup}

\noindent\textbf{Datasets.}
We trained every model with \textit{warm-start} method \cite{cooper2019zero}. We used LJSpeech~\cite{ito2017lj} for the pre-training phase and VCTK~\cite{veaux2016superseded} for the multi-speaker training phase.
Three VCTK speakers were entirely excluded due to their missing or inadequate data.
To evaluate objective metrics on unseen speakers, we held out eight VCTK speakers (four men and four women, each of whom has about 400 utterances) during training.
We also split the data of the remaining 98 VCTK speakers into training and validation sets for seen speaker evaluation, and the validation set had mostly 10 utterances for each speaker.

\noindent\textbf{Pre-processing.}
We used a character sequence as an input to the Tacotron encoder, where each character is represented as a 512-dim character embedding.
We downsampled an audio to 16kHz, and trimmed leading and trailing silence using librosa~\cite{mcfee2015librosa}.
We generated a spectrogram by 2048 point Fourier transform with Hann windowing with 16 ms shift and 64 ms length.
Finally, we converted it to a mel spectrogram with 80 frequency bins spanning from 125 Hz to 7.6 kHz.

\noindent\textbf{Implementation.}
In the coarse-grained and fine-grained encoders, the convolutional layers have 512 channels with $3\times1$ kernels, and the bidirectional LSTM layers have 256 cells for each direction.
The sizes of the global embedding, $d_g$, and the variable-length embedding, $d_v$, are both 256. 
The dimension of the attention key and query in the fine-grained encoder, $d_m$, is 256.
We follow the details described in \cite{shen2018natural} to implement the Tactoron 2 backbone.

\noindent\textbf{Training.}
We used the Adam optimizer \cite{kingma2014adam} with $\beta_1 = 0.9$, $\beta_2 = 0.999$, $\epsilon = 10^{-8}$, and a weight decay value of $10^{-6}$.
We minimized reconstruction loss, as described in \cite{shen2018natural}.
We pre-trained each model on LJSpeech data for 30k steps with the initial learning rate of $10^{-3}$ decaying to $10^{-4}$ at 20k steps.
Then, we resumed training the models on VCTK data for 70k steps with the learning rate of $10^{-4}$ decaying to $10^{-5}$ at 50k steps.
The batch size was 16 for the whole training process.

\subsection{Evaluation Method}

\noindent\textbf{Models.}
We refer to our proposed model as \textit{Attentron} and baselines, which can synthesize speech from speakers unseen during training as \textit{LDE}~\cite{cooper2019zero} and \textit{GMVAE}~\cite{hsu2018hierarchical}.

\begin{itemize}[leftmargin=*]
    \item \textbf{\textit{LDE}(a)} transfers well-trained speaker embedding space into multi-speaker TTS.
    We utilized the official implementation of the speaker verification model to extract speaker embeddings for multi-speaker Tacotron.
    The parenthesized parameter, \textit{a}, is the number of samples used to extract a speaker embedding.

    \item \textbf{\textit{GMVAE}(a)} is a controllable TTS model based on the variational autoencoder (VAE) framework.
    Since an official implementation is absent, we made an honest attempt to reproduce the results.
    To stabilize the training process, we adopted KL annealing as in \cite{zhang2019learning}.
    The parenthesized parameter, \textit{a}, is the number of samples fed into the encoder during inference.
    
    \item \textbf{\textit{Attentron}(a-b)}.
    The first parameter in the parenthesis, \textit{a}, represents the number of samples fed into the fine-grained encoder during training. 
    The latter parameter, \textit{b}, corresponds to the number of the samples fed into both the fine-grained encoder and the coarse-grained encoder during inference.  
\end{itemize}

\noindent\textbf{Metrics.}
We utilize the following metrics to evaluate models:

\begin{itemize}[leftmargin=*]
    \item \textbf{MCD-DTW} compares compatibility between the spectra of two audio sequences.
    Since sequences are not aligned, dynamic time warping is performed prior to comparison~\cite{kubichek1993mel}.

    \item \textbf{Speaker similarity} evaluates how much the synthesized speech resembles the target speaker.
    We extracted x-vectors from the synthesized speech as well as the actual speech of the target speaker and measured cosine similarity between them using the deep learning package named Resemblyzer~\cite{wan2018generalized}. 
    The value ranges from 0 to 1.

    \item \textbf{Attention collapse count} adds up significant intelligibility errors~\cite{he2019robust}.
    Without well-formed decoder attention, the output sounds incomprehensible and never ends on time.
    We report an attention collapse when the number of the generated spectrogram frames exceeds the pre-defined threshold ($4 \times input\_text\_length$ in our experiment). 
\end{itemize}

\noindent\textbf{User study.}
We performed user tests to evaluate human preference as to \textit{Naturalness} and \textit{Similarity}.
We used the mean opinion score (MOS) to rate user preference on a scale of 1-5, 5 being the best.
We utilized the subjective MOS evaluation method proposed by Jia \textit{et al.}~\cite{jia2018transfer}, which pairs each synthesized speech with the reference speech, to evaluate similarity.
We used Amazon's Mechanical Turk to collect subjective evaluations from crowd-sourced native speakers.
We selected eight seen speakers and unseen speakers (four men and women each) from VCTK data, and randomly selected 10 fixed examples from evaluation data.
The samples were presented at a fixed frame rate of 16kHz and at least 20 evaluators participated in each experiment.

\section{Experimental Result}\label{section:result}

\subsection{Evaluation Result}

\begin{table*}[t]
    \caption{
    Evaluation result. MCD-DTW (MCD), speaker similarity (Sim), and attention collapse count (Fail) are the objective metrics acquired from 960 and 3222 utterances of seen speakers and unseen speakers, respectively. 
    The subjective metrics, naturalness MOS (Nat-MOS) and similarity MOS (Sim-MOS), are presented with 95\% confidence intervals.
    Upward/downward pointing arrows correspond to metrics that are better when the values are higher/lower. 
    Bold values correspond to the best values of each metric.
    }
    \centering
    {\setlength{\tabcolsep}{0.66em}
    \begin{tabular}{c|ccccc|ccccc}
        \toprule
        \multirow{2}{*}{Model} & \multicolumn{5}{c|}{Seen speaker} & \multicolumn{5}{c}{Unseen speaker} \\
        & MCD$\downarrow$ & Sim$\uparrow$ & Fail$\downarrow$ & Nat-MOS$\uparrow$ & Sim-MOS$\uparrow$ & MCD$\downarrow$ & Sim$\uparrow$ & Fail$\downarrow$ & Nat-MOS$\uparrow$ & Sim-MOS$\uparrow$ \\
        \midrule
        Groundtruth & - & - & - & 4.13 $\pm$ 0.04 & 4.81 $\pm$ 0.02                & - & - & - & 4.15 $\pm$ 0.04 & 4.83 $\pm$ 0.02 \\
        \midrule
        LDE(1) & 12.85 & 0.731 & 43 & 3.56 $\pm$ 0.06 & 3.11 $\pm$ 0.07            & 14.38 & 0.677 & 82 & 3.75 $\pm$ 0.05 & 2.88 $\pm$ 0.07 \\
        GMVAE(1) & 12.51 & 0.774 & 23 & 3.61 $\pm$ 0.05 & 3.22 $\pm$ 0.06         & 13.94 & 0.686 & 59 & 3.76 $\pm$ 0.05 & 3.17 $\pm$ 0.06 \\
        Attentron(1-1) & \textbf{12.25} & \textbf{0.784} & \textbf{9} & \textbf{3.63} $\pm$ 0.05 & \textbf{3.33} $\pm$ 0.06    & \textbf{13.20} & \textbf{0.731} & \textbf{25} & \textbf{3.86} $\pm$ 0.05 & \textbf{3.30} $\pm$ 0.06 \\
        \midrule
        LDE(8) & 11.49 & 0.796 & 24 & 3.73 $\pm$ 0.05 & 3.48 $\pm$ 0.06      & 13.50 & 0.709 & 39 & 3.91 $\pm$ 0.05 & 3.17 $\pm$ 0.06 \\
        GMVAE(8) & 11.34 & 0.798 & \textbf{0} & 3.72 $\pm$ 0.05 & 3.40 $\pm$ 0.05           & 13.11 & 0.698 & 1 & 3.88 $\pm$ 0.04 & 3.27 $\pm$ 0.06 \\
        Attentron(8-8) & \textbf{10.99} & \textbf{0.812} & \textbf{0} & \textbf{3.76} $\pm$ 0.05 & \textbf{3.60} $\pm$ 0.05    & \textbf{11.67} & \textbf{0.788} & \textbf{0} &  \textbf{3.97} $\pm$ 0.04 & \textbf{3.57} $\pm$ 0.05 \\
        \bottomrule
    \end{tabular}
    }
    \vspace{-0.3cm}
    \label{tab:main_result_metric}
\end{table*}

Table \ref{tab:main_result_metric} shows the objective evaluation results, such as MCD-DTW, speaker similarity, and attention collapse count.
While there is no remarkable difference for seen speakers, proposed models significantly outperform the baseline models for the synthesis targeting unseen speakers.
Since it is not suggested how to make use of multiple reference audio files in the GMVAE model~\cite{hsu2018hierarchical}, we apply the $N$-shot inference method to it for a fair comparison by averaging out the embeddings.

The large improvement in the unseen speaker similarity scores, 7.9\% and 6.6\% comparing \textit{Attentron}(1-1) to \textit{LDE}(1) and \textit{GMVAE}(1), respectively, denotes that the proposed model clones a target unseen speaker's voice much closer than the others.
In addition, the fewer attention collapse counts of \textit{Attentron}(1-1), i.e., 9, show that our model outperforms baselines in the speech quality.
The baseline occasionally fails to form a proper attention alignment resulting in an unintelligible output speech, which is probably due to the absence of coarse-grained encoder in \textit{LDE}(1) and training difficulty of VAE in \textit{GMVAE}(1).

Furthermore, the strength of the proposed model to clone unseen speakers is more evident while utilizing eight reference audio files as an input.
\textit{Attentron}(8-8) significantly improves MCD-DTW and speaker similarity of unseen speakers, 11.6\% and 7.9\%, respectively, compared to \textit{Attentron}(1-1), whereas the baselines improve slightly using eight samples.
This result tells us that our model benefits from using multiple reference inputs more effectively than the baselines.

\subsection{User Study}
The MOS of naturalness and speaker similarity are listed in Table \ref{tab:main_result_metric}.
The proposed model slightly improves the naturalness compared to the baselines under the condition utilizing the identical number of reference samples.
The subjective metrics are consistent with the objective metrics showing the proposed method achieves a substantial improvement in the unseen speaker similarity without sacrificing the naturalness of speech.
We find that the naturalness MOS on unseen speakers is higher than seen speakers, by approximately 0.2 points on every model. 
We conjecture that it is because the naturalness scored by users varies with randomly sampled target speakers in the dataset that have different recording environments and speaking styles.

\subsection{Ablation Study}\label{subsection:ablation}

\begin{table}[t]
    \centering
    \caption{Ablation study for verifying impact of utilizing multiple reference inputs, coarse-grained encoder, fine-grained encoder and leveraging feature near raw reference audio.}
    \begin{tabular}{c|cc|cc}
        \toprule
        \multirow{2}{*}{Model} & \multicolumn{2}{c|}{Seen speaker} & \multicolumn{2}{c}{Unseen speaker} \\
        & MCD$\downarrow$ & Sim$\uparrow$ & MCD$\downarrow$ & Sim$\uparrow$ \\
        \midrule
        Attentron(8-8) & 10.99 & 0.812 &  11.67 & 0.788 \\
        Attentron(8-1) & 12.94 & 0.769     & 13.26 & 0.753 \\
        Attentron(1-8) & 11.04 & 0.809     & 12.19 & 0.749 \\
        \midrule
        w/o CE & 12.64 & 0.748 &  13.47 & 0.738 \\
        \midrule
        w/o FE & 10.98 &	0.819 &  12.20 & 0.757 \\
        Average pooling & 10.71 & 0.817 &  11.88 & 0.751 \\
        Self-attention & 10.62	& 0.821 &  11.86 & 0.755 \\
        \midrule
        Encoded value& 10.76 & 0.819 &  11.96 & 0.754 \\
        \bottomrule
    \end{tabular}
    \vspace{-0.3cm}
    \label{tab:ablation_study}
\end{table}

Table \ref{tab:ablation_study} shows the result of the ablation study.
We evaluate key components of the proposed model by tweaking \textit{Attentron}(8-8).
First, we further analyze the impact of multiple reference inputs.
Either utilizing multiple inputs during training (denoted as \textit{Attentron}(8-1)) or inference (denoted as \textit{Attentron}(1-8)) improves unseen speaker similarity compared to utilizing single reference input.
In addition, the result of \textit{Attentron}(8-8) is better than that of \textit{Attentron}(1-8) or \textit{Attentron}(8-1).
It suggests that the proposed model maximizes benefits from using multiple reference inputs by utilizing them during both training and inference.

We address the impact of the coarse-grained encoder.
Without coarse-grained encoder (denoted as \textit{w/o CE}), it loses its stability to generate an intelligible speech accompanying 35 and 143 attention collapse counts for seen speakers and unseen speakers, respectively.
Consequently, it shows worse MCD-DTW and speaker similarity than \textit{Attentron}(8-8).
Note that the attention collapse column is omitted from the table for clarity.

Next, we examine the fine-grained encoder in more detail.
To verify the effectiveness of the variable-length embedding, we remove the fine-grained encoder (denoted as \textit{w/o FE}) or use a single embedding.
The single embedding is obtained by replacing the attention module with an average pooling (denoted as \textit{Average pooling}) or a self-attention (denoted as \textit{Self-attention})~\cite{arik2018neural}.
These three cases make it hard to extract sufficient information from the reference audio files to clone unseen speakers, and thus the unseen speaker metrics are inferior.

We also test encoding the attention value by replacing one fully-connected layer into two convolutional layers, followed by two bidirectional LSTM layers (denoted as \textit{Encoded value}).
The seen speaker similarity of \textit{Encoded value} and \textit{Attentron}(8-8) are comparable.
The unseen speaker similarity of \textit{Encoded value}, however, significantly decreases compared to \textit{Attentron}(8-8), from 0.788 to 0.754.
We consider that manipulating the reference audio makes the model vulnerable to overfitting to seen speakers, which causes a large decline in the unseen speaker similarity.
It may induce the model to memorize the voice during training rather than imitate it on-the-fly.

\section{Related Works} \label{section:relatedwork}

Previous studies led to decent results for few-shot TTS, which, however, require an additional fine-tuning process~\cite{kons2019high, chen2018sample, bollepalli2019lombard, deng2018modeling, arik2018neural}. 
To avoid it, some approaches~\cite{Park2019MultiSpeakerES, nachmani2018fitting, arik2018neural, hu2019neural, cooper2019zero, jia2018transfer, chen2019cross} jointly or individually trained a speaker encoder generating a global embedding and utilized it to condition the TTS model.
However, conditioning the TTS model using only the global embedding is not sufficient to clone unseen speakers.

Similar to our work, some works made use of a variable-length embedding maintaining temporal information~\cite{lee2019robust, sun2020generating}.
Lee and Kim~\cite{lee2019robust} introduced an attention module to obtain a variable-length embedding.
However, it mainly focused on fine-grained control of prosody, and cannot clone unseen speakers since it utilizes a speaker lookup table which supports only seen speakers.
Sun \textit{et al.}~\cite{sun2020generating} introduced a fine-grained VAE model.
It also aimed to provide finer-level interpretations of prosody control and suggested autoregressive prior, differing from our interests in cloning unseen speakers with a few samples.

\section{Conclusion} \label{section:conclusion}

We proposed the novel architecture of multi-speaker TTS for cloning unseen speakers with a few samples.
It exploits two types of embeddings, the variable-length and global embedding generated by the fine-grained and coarse-grained encoder, respectively.
The fine-grained encoder is in particular designed to extract proper characteristics from relevant positions in an arbitrary number of reference audio samples.
By these means, it achieved high-quality synthesized speech of unseen speakers.
Our experimental results, including human evaluation, showed its excellence in terms of speaker similarity and speech quality.

It would be interesting to explore methods to control the prosody of the proposed few-shot TTS system, and we leave it for future work.
Another avenue of future work is cloning unseen speakers with a few samples in a voice conversion task.

\bibliographystyle{IEEEtran}
\bibliography{mybib}

\end{document}